\documentclass{ws-ijmpa-blank-new}

\usepackage[super,compress]{cite}
\usepackage{graphicx}

\usepackage{amssymb,epsf,latexsym,amssymb,amsmath,mathrsfs}
\usepackage[english]{babel}
\usepackage{dsfont}  

\newcommand{\beq}{\begin{equation}}
\newcommand{\eeq}{\end{equation}}
\newcommand{\beqa}{\begin{eqnarray}}
\newcommand{\eeqa}{\end{eqnarray}}
\newcommand{\bsubeqs}{\begin{subequations}}
\newcommand{\esubeqs}{\end{subequations}}

\newcommand{\christoffel}[3]{\Gamma^{#1}_{\hphantom{#1}#2#3}}

\makeatletter
\@addtoreset{equation}{section}
\renewcommand{\theequation}{\thesection.\arabic{equation}}

\makeatother

\begin{document}
\markboth{F.R. Klinkhamer and Z.L. Wang}
{Lensing and imaging by a stealth defect of spacetime}

%
\catchline{}{}{}{}{}
%

\title{\vspace*{-11mm}
Lensing and imaging by a stealth defect of spacetime}

\author{F.R. Klinkhamer}
\address{Institute for Theoretical Physics, Karlsruhe Institute of
Technology (KIT),\\ 76128 Karlsruhe, Germany\\
frans.klinkhamer@kit.edu}

\author{Z.L. Wang}
\address{Institute for Theoretical Physics, Karlsruhe Institute of
Technology (KIT),\\ 76128 Karlsruhe, Germany\\
ziliang.wang@kit.edu}

\maketitle


\begin{abstract}
We obtain the geodesics for the simplest possible
stealth defect which has a flat spacetime.
We, then, discuss the lensing properties of such a defect,
and the corresponding image formation.
Similar lensing properties can be expected to hold
for curved-spacetime stealth defects.
\end{abstract}
\vspace*{.0\baselineskip}
{\footnotesize
\vspace*{.25\baselineskip}
\noindent \hspace*{5mm}
\emph{Journal}: Mod. Phys. Lett. A \textbf{34} (2019) 1950026
\vspace*{.25\baselineskip}
\newline
\hspace*{5mm}
\emph{Preprint}: arXiv:1808.02465  
}
\vspace*{-5mm}\newline
\ccode{PACS Nos.: 04.20.Cv, 04.20.Gz, 98.62.Sb}


\section{Introduction}
\label{sec:Introduction}

A particular Skyrmion spacetime defect has been studied
recently in a series of papers:
the self-consistent \emph{Anz\"{a}tze} for the fields
were established in Ref.~\citen{Klinkhamer2014-prd},
the origin of a possible negative asymptotic
gravitational mass was discussed in
Ref.~\citen{KlinkhamerQueiruga2018-antigrav}, and the details
of a special defect solution with zero asymptotic gravitational
mass were given in Ref.~\citen{KlinkhamerQueiruga2018-stealth}.
This last defect solution, with
positive energy density of the matter fields
but a vanishing asymptotic gravitational mass,
has been called a ``stealth defect.''

It was stated in Sec.~4 of Ref.~\citen{KlinkhamerQueiruga2018-stealth}
that ``assuming the existence of this particular type of
spacetime-defect solution without long-range fields,
an observer has no advance warning if he/she approaches
such a stealth-type defect solution
(displacement effects of background stars are negligible,
at least initially).''
The goal of the present article is to expand on the
parenthetical remark of the previous quote.
We study, in particular, the geodesics of the
special defect solution of
Fig.~5 in Ref.~\citen{KlinkhamerQueiruga2018-stealth},
which has a flat spacetime.
For completeness, we have also performed
a simplified calculation of the geodesics
for a curved-spacetime stealth defect and present the results
in \ref{app:Geodesics-curved-spacetime-stealth-defect}.

It may be helpful to place the present paper in context
before we start our somewhat technical discussion.
The important point to realize is that the spacetime
manifolds discussed in
Refs.~\citen{Klinkhamer2014-prd,KlinkhamerQueiruga2018-antigrav,%
KlinkhamerQueiruga2018-stealth}
are genuine solutions of the standard Einstein equation,
but with a nontrivial spacetime topology
and a degenerate metric
(regarding this last characteristic,
see, in particular, the second and third remarks
in Sec. VI of Ref.~\citen{KlinkhamerQueiruga2018-antigrav}).
The analysis of the main part of the present paper is
for an  \emph{exact} solution of the vacuum Einstein equation,
namely the flat-spacetime defect solution.
The analysis in the Appendix  of the present paper
is for an \emph{approximation}
of the numerical solution
of the Einstein and matter-field equations.
The results in the Appendix are, therefore, only indicative.

Another point that needs to be clarified in advance
is the order of magnitude of the defect length scale $b$,
as defined in Sec.~\ref{sec:Geodesic equations}
and the caption of Fig.~\ref{fig1}.   
In Sec.~\ref{sec:Discussion},
we will briefly discuss a ``gas'' of identical static
defects. In that case, the experimental data are
consistent with having a highly-rarified gas of
microscopic  static defects (e.g., $b \sim l_\text{planck}
\equiv \sqrt{8\pi G_{N}\, \hbar/c^3}
\approx 8.10 \times 10^{-35}\;\text{m}$ and
a typical distance between the individual defects of order $l \gg b$).
Still, nothing excludes having, in a remote part of the Universe,
a single spacetime defect with a macroscopic value of its length scale
(e.g., $b\sim \text{Mpc}\approx 3.09 \times 10^{22}\;\text{m} $).

\section{Geodesic equations}
\label{sec:Geodesic equations}

The topology and coordinatization of the spacetime manifold
considered has been reviewed in Sec.~2 of Ref.~\citen{Klinkhamer2014-mpla}
and Sec.~II D of Ref.~\citen{KlinkhamerSorba2014}
(see also Chap.~3 of Ref.~\citen{Guenther2017}    
for the proper definition of the field equations).
Very briefly, the spatial part of the manifold is obtained by removing
the interior of a ball in three-dimensional Euclidean space
and by identifying antipodal points on the boundary of this ball
(the defect surface has topology
$\mathbb{R}P^2 \sim S^2/\mathbb{Z}_2$).
As to the coordinatization, there are three coordinate charts.
Here, we focus on the chart-2 coordinates,
the other charts being similar.
Moreover, we use dimensionless coordinates, all lengths
being measured in units of $1/(e\,f)>0$ for the theory as
defined in Ref.~\citen{KlinkhamerQueiruga2018-stealth}.

The metric of a particular defect-type solution of the vacuum Einstein equation
reads as follows (cf. Sec.~3 of Ref.~\citen{Klinkhamer2014-mpla}):
\bsubeqs\label{eq:vacuum-metric-w-def-y0-def}
\beqa\label{eq:vacuum-metric-general-l}
\hspace*{-5mm}
ds^{2}\,\Big|^\text{(vac.\,sol.)}
&=&
-\left(1-l/\sqrt{w}\,\right)\,(dt)^{2}
+\frac{1-y_{0}^{2}/w}{1-l/\sqrt{w}}\;(dy)^{2}
\nonumber\\
&&
+w\,\Big[(dz)^{2}+\sin^{2}z\,(dx)^{2}\Big]\,,
\\[2mm]
\label{eq:w-def}
\hspace*{-5mm}
w &\equiv& y_{0}^{2} +y^{2}\,,
\\[2mm]
\label{eq:y0-def}
\hspace*{-5mm}
y_{0} &\equiv& e\, f\, b  > 0\,,
\eeqa\esubeqs
where $y_{0}$ corresponds to the dimensionless version of
the defect length scale $b>0$.
Note that we only show the dimensionless chart-2 coordinates.
Specifically, the spatial chart-2 coordinates
have the following ranges:
\bsubeqs\label{eq:chart-coord-ranges}
\beqa
x &\in& (0,\pi)\,,
\\[2mm]
y &\in& (-\infty,\infty)\,,
\\[2mm]
z &\in& (0,\pi)\,,
\eeqa\esubeqs
where $x$ and $z$ are angular coordinates
and $y$ is a dimensionless quasi-radial coordinate
with $y=0$ corresponding to the defect surface
($y$ is positive one side of the defect and negative on the other).

For a globally regular solution, the real constant $l$ in
\eqref{eq:vacuum-metric-general-l} takes the following values:
\begin{equation}\label{eq:l-domain}
l \in (-\infty,\, y_{0})\,.
\end{equation}
With $l=0$ for the stealth-defect solution from
Sec.~2.4 and Fig.~5 in Ref.~\citen{KlinkhamerQueiruga2018-stealth},
we have the metric
\bsubeqs\label{eq:vacuum-metric-zero-l}
\begin{equation}
ds^{2}\,\Big|^{(\text{vac.\,sol.}\;l=0)}
=-(dt)^{2}+A(y)\;(dy)^{2}+w\,\Big[(dz)^{2}+\sin^{2}z\,(dx)^{2}\Big]\,,
\end{equation}
with
\begin{equation}\label{eq:A-def}
A(y) =  \frac{y^{2}}{y_{0}^{2} +y^{2}}\,
\end{equation}
\esubeqs
and $w$ defined by \eqref{eq:w-def}.
Then, the nonvanishing Christoffel symbols are
\bsubeqs\label{eq:christoffel}
\beqa
\christoffel{y}{y}{y}&=&\frac{A'}{2A} \,,  \\[2mm]
\christoffel{y}{z}{z}&=&-\frac{w'}{2A} \,,   \\[2mm]
\christoffel{y}{x}{x}&=&-\frac{w'\sin ^{2} z}{2A}\,,  \\[2mm]
\christoffel{z}{y}{z}&=&\christoffel{z}{z}{y}=\frac{w'}{2w}\,,  \\[2mm]
\christoffel{z}{x}{x}&=&-\sin z \cos z \,, \\[2mm]
\christoffel{x}{y}{x}&=&\christoffel{x}{x}{y}=\frac{w'}{2w}\,,  \\[2mm]
\christoffel{x}{z}{x}&=&\christoffel{x}{x}{z}=\cot z \,,
\eeqa\esubeqs
where the prime stands for differentiation   
with respect to $y$.
The first three Christoffel symbols are divergent at the defect surface,
but our results will show that the motion of a particle can still be regular.

From the geodesic equation~\cite{Weinberg1972}
with affine parameter $\lambda$, we find
\bsubeqs\beqa
0 &=& \frac{d^{2}t}{d\lambda ^{2}}  \,,\label{eq:geodesic t}
\\[2mm]
0 &=& \frac{d^{2} y}{d \lambda ^{2}}+\christoffel{y}{y}{y}
\left(\frac{dy}{d\lambda}\right)^{2}+\christoffel{y}{z}{z}
\left(\frac{dz}{d\lambda}\right)^{2}+\christoffel{y}{x}{x}
\left(\frac{dx}{d\lambda}\right)^{2}  \,,\label{eq:geodesic y}
\\[2mm]
0 &=& \frac{d^{2} z}{d \lambda ^{2}}+2\christoffel{z}{y}{z}
 \frac{dy}{d\lambda}\frac{dz}{d\lambda}+\christoffel{z}{x}{x}
 \left(\frac{dx}{d\lambda}\right)^{2} \,,\label{eq:geodesic z}
\\[2mm]
0 &=& \frac{d^{2} x}{d \lambda ^{2}}+2\christoffel{x}{y}{x}
 \frac{dy}{d\lambda}\frac{dx}{d\lambda}+2\christoffel{x}{z}{x}
 \frac{dx}{d\lambda}\frac{dz}{d\lambda} \,.\label{eq:geodesic x}
\eeqa\esubeqs
We can choose the normalization of $\lambda$ so that the solution of~\eqref{eq:geodesic t} has
\begin{equation}\label{eq:solution t}
\frac{dt}{d\lambda}=1\,.
\end{equation}
Then, $\lambda$ can be replaced by $t$
in \eqref{eq:geodesic y}, \eqref{eq:geodesic z}, and \eqref{eq:geodesic x}.
Since the metric is spherically symmetric,
we need only consider the case $z=\pi/2$.
Our calculation follows Sec.~8.4
of Ref.~\citen{Weinberg1972}.

For the case $dx/dt\neq 0$, divide \eqref{eq:geodesic x}
by $dx/dt$ and use the Christoffel symbols from \eqref{eq:christoffel}.
We, then, have
\begin{equation}\label{eq:conservation-eq}
\frac{d}{•dt}\left(\ln \frac{dx}{dt}+\ln w \right)=0\,,
\end{equation}
which gives a real constant (up to a sign),
\begin{equation}\label{eq:definition-J}
J\equiv w \, \frac{dx}{dt}\,.
\end{equation}
With \eqref{eq:christoffel}, \eqref{eq:definition-J},
and multiplying \eqref{eq:geodesic y} by $2A\, dy/dt$, we find
 \begin{equation}
 \frac{d}{dt}\left[ A\left( \frac{dy}{dt}\right) ^{2}+\frac{J^{2}}{w} \right]=0\,.
 \end{equation}
Hence, there is the following constant of motion:
\begin{equation}\label{eq:definition E}
E\equiv A\left(\frac{dy}{dt}\right)^{2}+\frac{J^{2}}{w}\,.
\end{equation}
By elimination of $t$ from \eqref{eq:definition-J} and \eqref{eq:definition E},
we get $y$ as a function of $x$,
\begin{equation}\label{eq:orbit}
\frac{A}{w^{2}}\left(\frac{dy}{dx}\right)^{2}+\frac{1}{w}=\frac{E}{J^{2}}\,.
\end{equation}
From \eqref{eq:definition-J}, \eqref{eq:definition E}, and $z=\pi/2$,
the metric \eqref{eq:vacuum-metric-zero-l}
along the geodesic
can now be written as
\begin{equation}
ds^{2}=(-1+E)(dt)^{2}\,.
\end{equation}
In other words, we have
\bsubeqs\label{eq:E-range}   
\beqa
\label{eq:E-range-massless}
E &=& 1\,,        \qquad\;\;\; \text{for a massless particle}\,,
\\[2mm]
\label{eq:E-range-massive}
E &\in& [0,\,1)\,, \quad       \text{for a massive particle}\,,
\eeqa
\esubeqs
where the case $E=0$ corresponds to $dy/dt=0$, as will be
discussed in Sec.~\ref{sec:Radial geodesics}.

\section{Radial geodesics}
\label{sec:Radial geodesics}

\begin{figure}[t]
  \centering
  \includegraphics[width=0.6\textwidth]{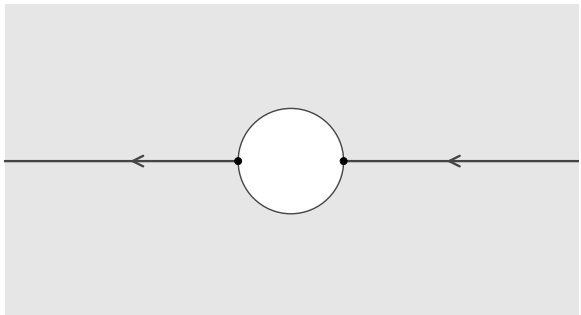}
  \caption{Radial geodesic
  for the stealth defect \eqref{eq:vacuum-metric-zero-l},
  with part of the 3-dimensional space manifold indicated by the shaded area
  and antipodal points (dots) on the defect surface identified.
  The ``long distance'' between the dots equals $\pi b$, where
  $b$ is the defect length scale.}
  \label{fig1}
\end{figure}

Consider the geodesic equation for a particle moving exactly
in the negative $y$ direction
(going from right to left in Fig.~\ref{fig1}), i.e., $dx/dt=0$.
From the definition of $J$ in \eqref{eq:definition-J},
it follows immediately that $J=0$,
even though $J$ was initially defined as a  nonzero
quantity [see the sentence at the start of the paragraph
above \eqref{eq:conservation-eq}].
The corresponding energy-type constant of motion is
\begin{equation}\label{eq:radial}
E= \frac{y^{2}}{y_{0}^{2}+y^{2}}\left(\frac{dy}{dt}\right)^{2}\,.
\end{equation}
The solutions of \eqref{eq:radial} are
\bsubeqs\label{eq:radial solution1}
\beqa
y &=& \pm \sqrt{-y_{0}^{2}+\left(+\sqrt{E}\,t+C_1\right)^{2}}\,,
\\[2mm]
y &=& \pm \sqrt{-y_{0}^{2}+\left(-\sqrt{E}\,t+C_2\right)^{2}}\,,
\eeqa
\esubeqs
where $C_1$ and $C_2$ are real constants.

Making appropriate time shifts (or setting  $C_1=C_2=y_{0}$)
and defining $B \equiv \sqrt{E}/y_{0}$,
the solutions \eqref{eq:radial solution1} reproduce
the results of Sec.~3 in Ref.~\citen{Klinkhamer2014-mpla}.
Finally, as mentioned below \eqref{eq:E-range-massive},
we find a constant $y$ solution if $E=0$.
%
%

\section{Nonradial geodesics}
\label{sec:Nonradial-geodesics}

Nonradial geodesics exist in two types,
those which cross the defect surface  
\mbox{($\mathbb{R}P^2 \sim S^2/\mathbb{Z}_2$)}
and those which do not.

\subsection{Geodesics not crossing the defect surface}
\label{subsec:Geodesics-not-crossing-the-defect-surface}

Outside the defect surface,
the spacetime \eqref{eq:vacuum-metric-zero-l}
is Minkowskian with vanishing curvature invariants~\cite{Klinkhamer2014-mpla}.
So, geodesics which do not cross the defect surface
should be straight lines with standard Cartesian coordinates.
The following calculation will show this explicitly.

\begin{figure}[t]
  \centering
  \includegraphics[width=0.6\textwidth]{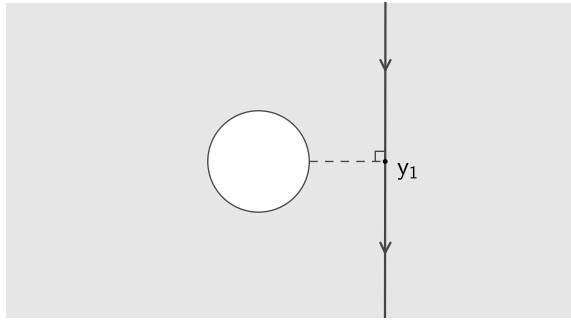}
  \caption{Nonradial geodesic which does not cross the defect surface
  and defines the quasi-radial coordinate $y_1>0$.}
  \label{fig2}
\end{figure}

From \eqref{eq:orbit}, we find
\begin{equation}\label{eq:orbit1}
dx=\pm \int \frac{y\,dy}{(y_{0}^{2}+y^{2})\sqrt{(E/J^{2})\,(y_{0}^{2}+y^{2})-1}}\,.
\end{equation}
Define the quasi-radial coordinate $y_1$ corresponding to
the point on the line closest to the defect surface
(cf. Fig.~\ref{fig2} with $y_1>0$),
so that $|y_1|$ corresponds to an ``impact parameter.''
Since $d\sqrt{w}/dx$ and $dy/dx$ vanish at $y_1$,
\eqref{eq:orbit} gives
\begin{equation}
\frac{1}{y_{0}^{2}+y_1^{2}}=\frac{E}{J^{2}}\,.
\end{equation}
Then, \eqref{eq:orbit1} can be written as
\begin{equation}\label{eq:x(y)}
x(y)=x(\infty)\pm \int^ \infty _y \frac{y\,dy}{(y_{0}^{2}+y^{2})
\sqrt{(y_{0}^{2}+y^{2})/(y_{0}^{2}+y_1^{2})-1}}\,.
\end{equation}
At $y=y_1$, \eqref{eq:x(y)} gives
\begin{equation}\label{eq:change-of-x}
|x(y_1)-x(\infty)|=\pi/2\,.
\end{equation}
The result \eqref{eq:change-of-x} shows that these particular geodesics
(nonradial and nonintersecting with the defect surface)
are indeed straight lines.

\subsection{Geodesics crossing the defect surface}
\label{subsec:Geodesics-crossing-the-defect-surface}

Now, consider nonradial geodesics which cross the defect surface.
If we use in \eqref{eq:orbit} the replacement
\begin{equation}\label{eq:replacement-dy/dx}
\frac{dy}{dx}=\frac{1}{2y}\,\frac{dy^{2}}{dx}\,,
\end{equation}
we find the following two solutions for $y^{2}$:
\bsubeqs\label{eq:solution_orbit}
\beqa
y^{2} &=& \frac{\tan^{2}(x_1+x)+1}{E/J^{2}}-y_{0}^{2} \,,
\\[2mm]
y^{2} &=& \frac{\tan^{2}(x_2-x)+1}{E/J^{2}}-y_{0}^{2} \,,
\eeqa
\esubeqs
where $x_1$ and $x_2$ are real constants.

Note that the metric~\eqref{eq:vacuum-metric-zero-l}
has a spherically symmetric form
and that the corresponding ``radial" coordinate is $\sqrt{w}\in [y_{0},\infty)$.
After a shift of the constants, the solutions~\eqref{eq:solution_orbit}
can be written as%
\bsubeqs\label{eq:solution-w}
\beqa
\sqrt{w}\,\sin(x_1-x) &=& \pm \frac{J}{\sqrt{E}} \,,
\\[2mm]
\sqrt{w}\,\sin(x_2+x) &=& \pm \frac{J}{\sqrt{E}} \,,
\eeqa
\esubeqs
with $\sqrt{w} \geq y_{0}$.
Several comments on the solutions \eqref{eq:solution-w}
are in order:  
\begin{enumerate}
\item[(i)]
mathematically, the solutions are straight lines or straight-line segments in polar-type coordinates ($\sqrt{w},\,x$);
\item[(ii)]
the solutions are regular at the defect surface, $\sqrt{w}=y_{0}\,$;
\item[(iii)]
to find the complete geodesic of a given particle among these solutions,
we must remember the antipodal identifications at the defect surface
$\sqrt{w}=y_{0}$.
\end{enumerate}

\begin{figure}[t]
  \centering
  \includegraphics[width=0.6\textwidth]{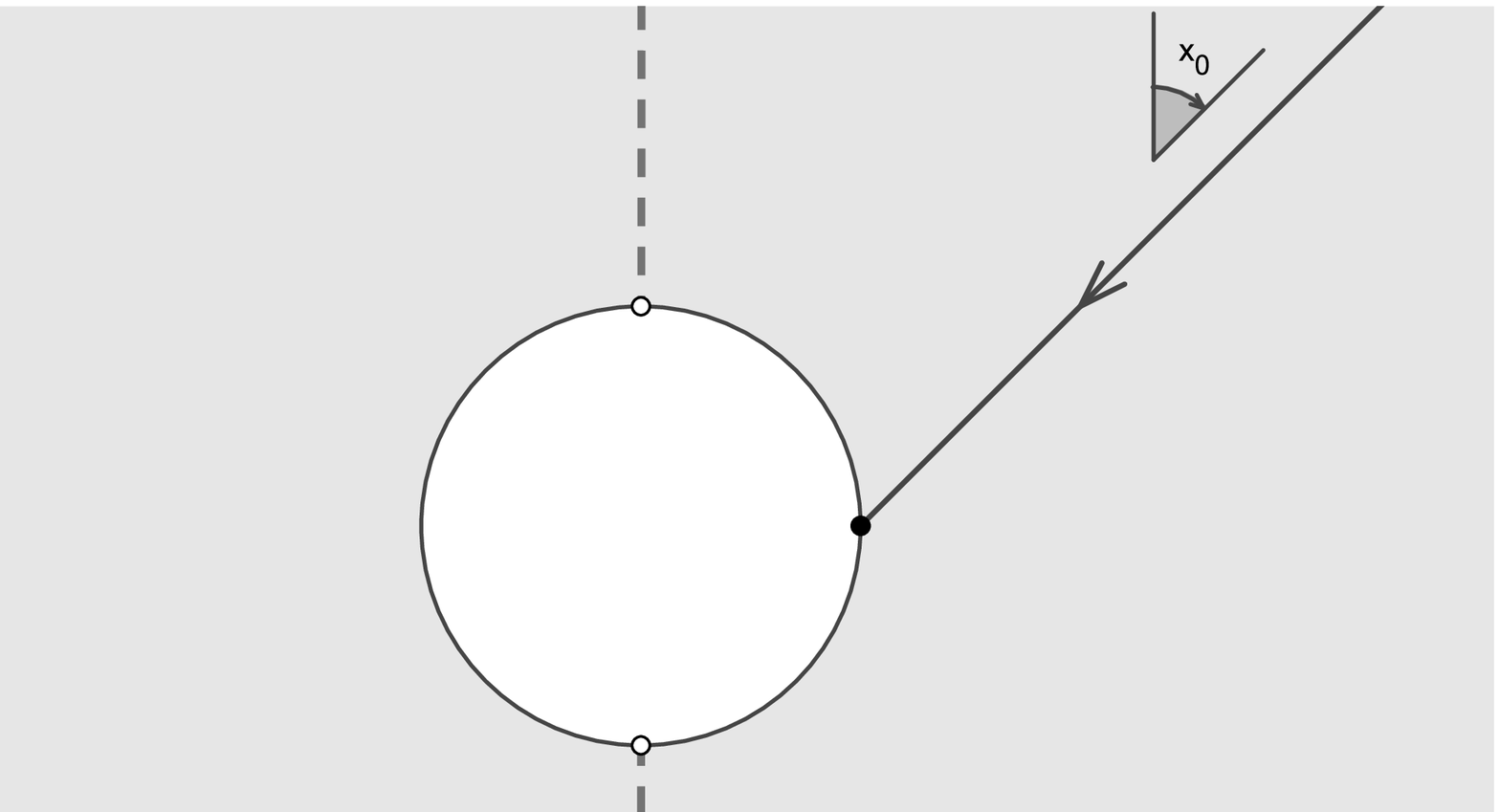}
  \caption{Ingoing line \eqref{eq:ingoing-line-formula}
  lying in the domain of the  chart-2 coordinates.
   The dashed line shows the $x_3$ Cartesian axis, which
   does not belong to the domain of the  chart-2
   coordinates~\cite{KlinkhamerSorba2014}.}
  \label{fig3}
\vspace*{.5cm}
  \centering
  \includegraphics[width=0.6\textwidth]{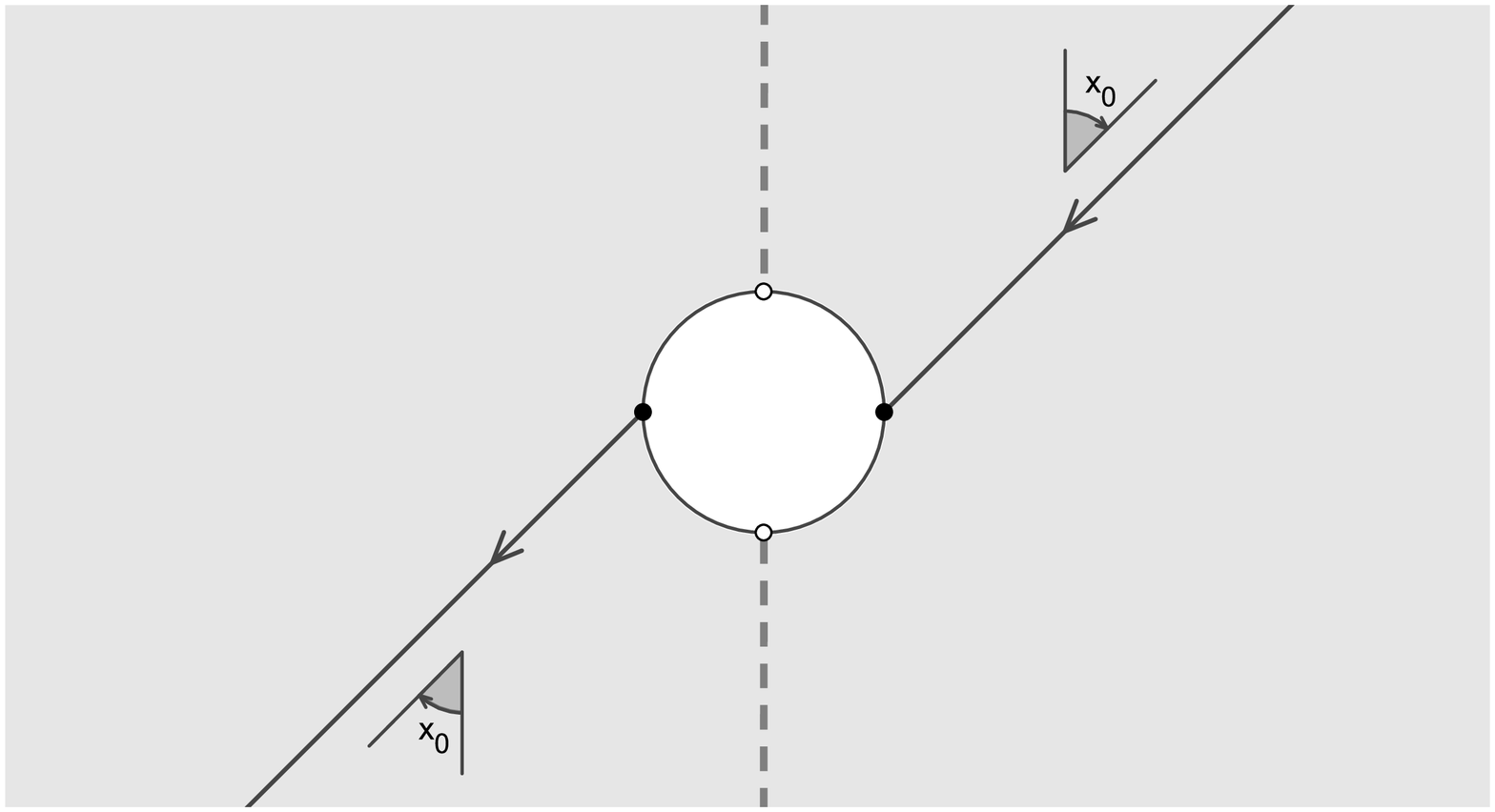}
  \caption{Nonradial geodesic crossing the defect surface.}
  \label{fig4}
\vspace*{.5cm}
  \centering
  \includegraphics[width=0.6\textwidth]{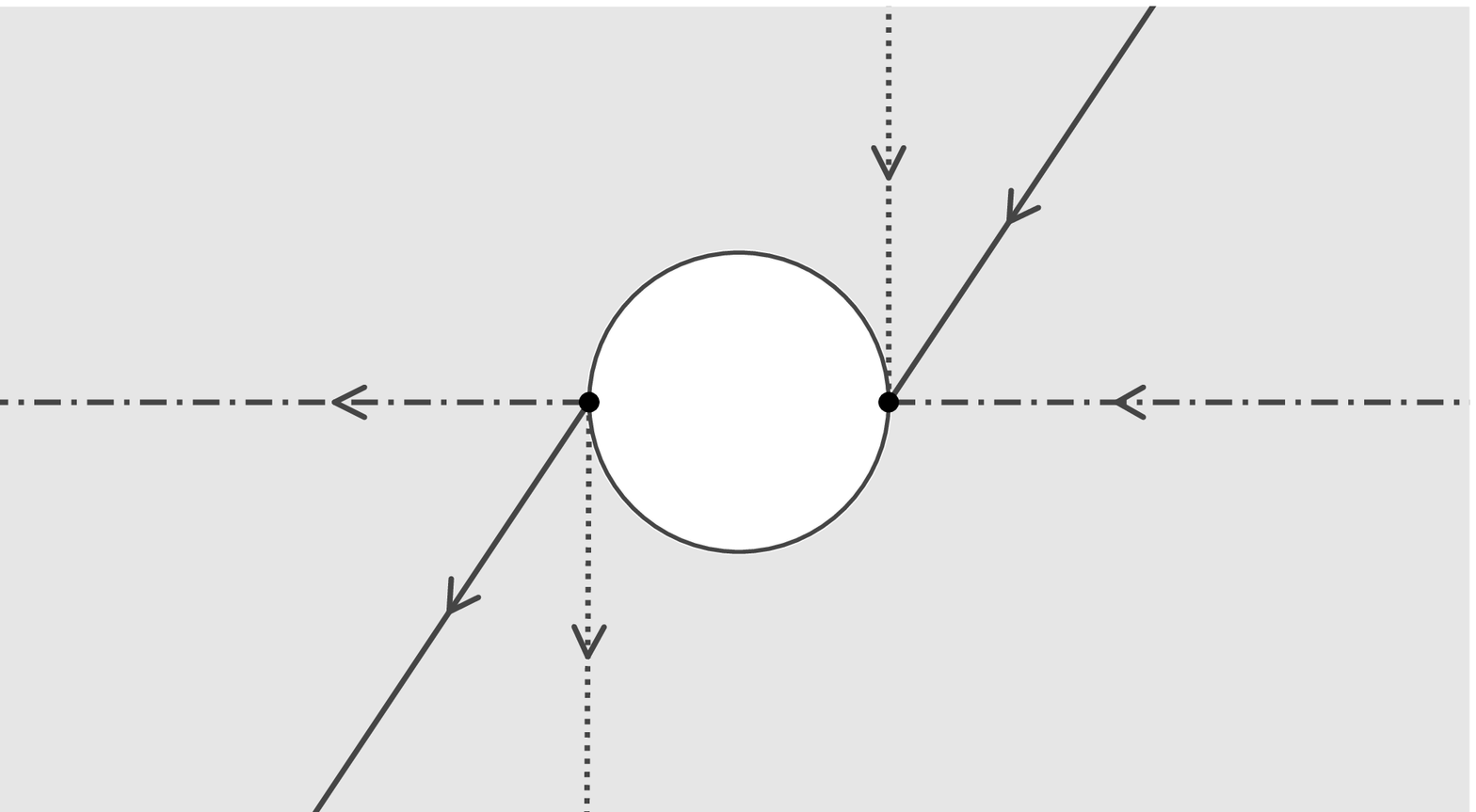}
  \caption{Family of geodesics crossing the defect surface.}
  \label{fig5}
\vspace*{0cm}
\end{figure}

For a nonradial ingoing line, it is convenient to choose coordinates,
so that the end of the ingoing line has $x=\pi/2$
(Fig.~\ref{fig3}).
In these coordinates, the ingoing line is given by
\bsubeqs\label{eq:ingoing-line-formula}
\beqa
\sqrt{w}\sin(x_{0}-x) &=& - y_{0} \cos x_{0} \,,
\eeqa
with
\beqa\label{eq:ingoing-line-formula-range}
0<x_{0}<x\leq \pi/2 \,.
\eeqa
\esubeqs
Observe that we have included the end point of the ingoing line
in \eqref{eq:ingoing-line-formula-range}.
We can check that the formula \eqref{eq:ingoing-line-formula}
indeed corresponds to one of the solutions \eqref{eq:solution-w}.

In this case, there will exist, among the solutions \eqref{eq:solution-w},
a unique outgoing line
(Fig.~\ref{fig4}) if the following two conditions are met:
\begin{enumerate}
\item
the beginning of the outgoing line and the end of the ingoing line
must be antipodal points at the defect surface (these points are
identified);
\item
the complete geodesic must be a straight line if $y_{0}=0$.
\end{enumerate}
Note that, with a nonradial ingoing line as in Fig.~\ref{fig4},
the quantity $J$ will change sign after crossing the defect surface
(see Ref.~\citen{KlinkhamerSorba2014}
for further discussion of the anomalous
angular-momentum behavior of scattering solutions).
Based on the above two points, Fig.~\ref{fig5} shows three geodesics
from a continuous family of geodesics crossing the defect surface:
the family ranges continuously from a radial geodesic (dot-dashed line)
to a tangent geodesic (dotted line).

From the particular family of geodesics as shown in Fig.~\ref{fig5},
we obtain what may be called a ``shifted tangent geodesic''  
(dotted line in Fig.~\ref{fig5}).
But, from the limiting case of the geodesic in Fig.~\ref{fig2}
with $y_1 \to  0^{+}$,
we obtain what may be called an ``ongoing tangent geodesic''
(solid line in Fig.~\ref{fig2} pushed towards the
defect surface).
Hence, we conclude that
``certain geodesics at the defect surface $y = 0$
cannot be continued uniquely,''
as mentioned in the second remark of
Sec.~VI in Ref.~\citen{KlinkhamerQueiruga2018-antigrav}
(further discussion can be found in Sec.~3.1.5
of Ref.~\citen{Guenther2017}).

\section{Image formation by a flat-spacetime stealth defect}
\label{sec:Image-formation}

The geodesics of the stealth-defect
spacetime \eqref{eq:vacuum-metric-zero-l}
have been discussed
in Secs. \ref{sec:Radial geodesics} and \ref{sec:Nonradial-geodesics}.
For a nonradial geodesic reaching the defect surface, Fig.~\ref{fig4}
shows that the defect causes
a parallel shift of the geodesic in the ambient space
(i.e., the Euclidean 3-space away from the defect surface).
In this section, we will show that this shift can, in principle,
create an image of a given object.

\begin{figure}[t]
  \centering
  \includegraphics[width=0.8\textwidth]{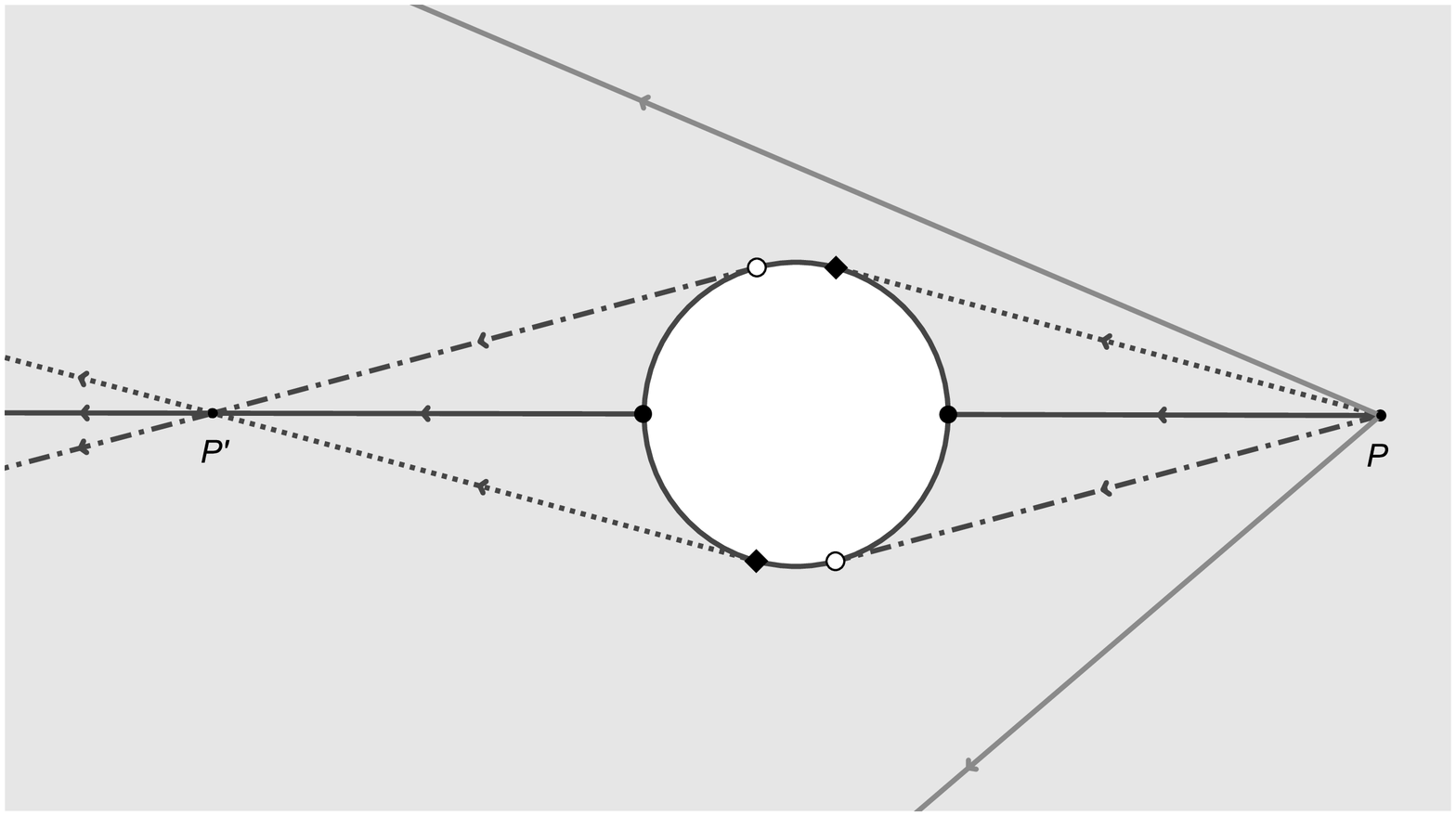}
  \caption{Geodesics with intersection points $P$ and $P'$.}
  \label{fig6}
\vspace*{.5cm}
  \centering
  \includegraphics[width=0.8\textwidth]{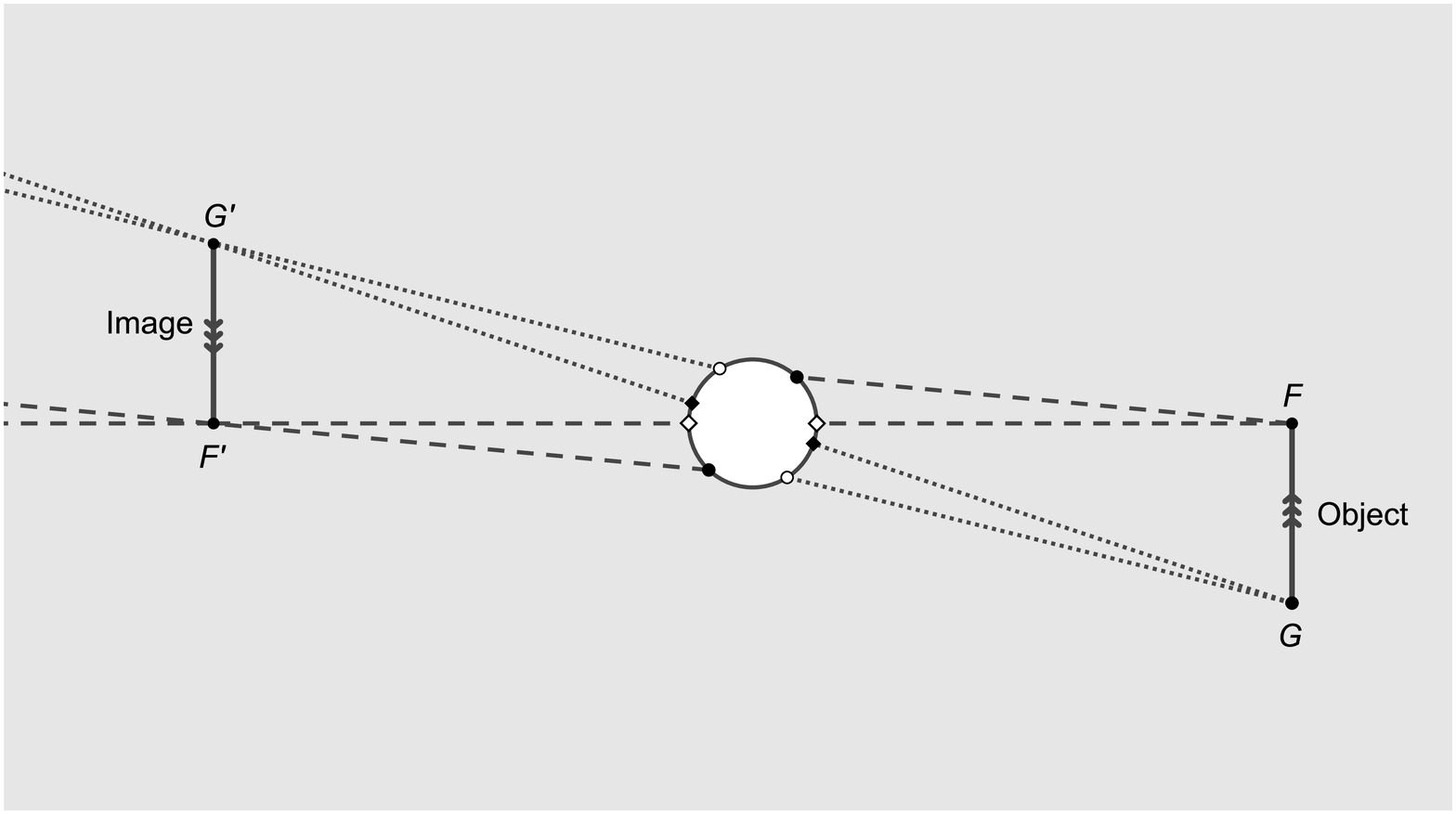}
  \caption{Image formation by a stealth defect.}
  \label{fig7}
\vspace*{0cm}
\end{figure}

First, consider geodesics which start from a point $P$ at one side
of the defect (Fig.~\ref{fig6}). For geodesics which cross the defect surface,
there will be an intersection point $P'$ at the other side of the defect.
In fact,
$P$ and $P'$ are reflection points about the ``center" of the defect
(considered to be obtained by surgery on the three-dimensional Euclidean space).
The different paths connecting $P$ and $P'$
have, in general, different values for the time-of-flight.

Next, observe that,
based on the above discussion for the geodesics of a stealth-defect spacetime,
a permanent luminous object will give a real image of the object (Fig.~\ref{fig7}).
The qualification ``permanent'' for the light source
refers to the different time-of-flight values mentioned in the
previous paragraph.

Several additional remarks are in order.
First, the image in Fig.~\ref{fig7} is located at the reflection
point on the other side of the defect.

Second, the image is inverted and the image size is equal to the object size.
Note that this is also the case if an object in Minkowski spacetime
is located at a $2f$ distance from a standard thin double-convex lens,
where $f$ is the focal length of the lens
(cf. Sec.~27.3 of Ref.~\citen{Feynman1963}).
Recall that the time-of-flight of different paths
connecting the $2f$ points of a standard lens in Minkowski spacetime
is equal, due to the reduced speed of light in the lens material and
the appropriate shape of the lens.

Third, if we consider the image
from a static luminous source, then the irradiance of the image
depends on both the defect scale $b$ and the location of the source
(the irradiance is defined as the power per unit receiving area;
cf. Secs.~5.3.2 and 5.3.5 of Ref.~\citen{MouroulisMacdonald1997}).  
The irradiance will be larger if $b$ is increased for an unchanged source position
(larger ``white disk'' in Fig.~\ref{fig7})
or if the source is brought closer to the defect for an unchanged defect scale
(object and image closer to the ``white disk'' in Fig.~\ref{fig7}):
in both cases, the flux captured and transmitted by the defect is larger.

Fourth, return to the analogy with standard lenses in Minkowski spacetime
as mentioned in the second remark and note that
our defect resembles a so-called zoom lens, with a finite range of focal lengths.
If we recall the standard lens equation
$1/d_\text{object}+ 1/d_\text{image}= 1/f$ in Minkowski spacetime
[cf. Eq.~(27.12) of Ref.~\citen{Feynman1963}],
we see that our defect has an effective focal length $f_\text{eff}$
given by
\beq
2\,f_\text{eff
}=\sqrt{b^2+(Y_\text{object})^2} \in (b,\, \infty)\,,
\eeq
where $Y_\text{object} \equiv y_\text{object}/(e f) \ne 0$ is the
dimensional chart-2 quasi-radial coordinate
of a small object away from the defect surface.

Fifth, if a permanent pointlike light source is
placed at point $P$ of Fig.~\ref{fig6},
then an observer at point $P'$ in the same figure will see a luminous disk
(different from the Einstein
ring~\cite{Einstein1936,VirbhadraEllis2000,Perlick2004,Schneider-etal1992,%
Wambsganss1998,Dodelson2017},
which the observer would see if the defect were replaced
by a patch of Minkowski spacetime with a static
spherical star at the center).

Sixth, it may be of interest to compare
our lensing and imaging results
from the spacetime defect with those from wormholes
(see, e.g., Refs.~\citen{Perlick2004,Shatskiy2004,%
Nandi-etal2006,ShaikhKar2017} and references therein).
In both cases, there is an unusual ingredient in the
physics setup: exotic matter for the wormholes
and a degenerate metric for the spacetime defect.

\section{Discussion}
\label{sec:Discussion}

In the present article, we have studied
the geodesics of the static $l=0$
stealth-defect solution  \eqref{eq:vacuum-metric-zero-l},
which has a flat spacetime.
This exact solution of the vacuum Einstein equation
results in Ref.~\citen{KlinkhamerQueiruga2018-stealth}
from the parameter choice
$\widetilde{\eta}\equiv 8\pi\, G_{N}\, f^2 = 0$
for $G_{N}=0$ and $f>0$.

Incidentally, \emph{exact} multi-defect solutions
of the vacuum Einstein equation are obtained by
superposition of these static $l=0$ defects,
as long as the individual defect surfaces do not intersect.
(Tight experimental bounds for such a Lorentz-violating
``gas'' of defects have been obtained in
Refs.~\citen{BernadotteKlinkhamer2007,KlinkhamerSchreck2008}.)
It may even be possible to obtain an exact
multi-defect solution of the vacuum Einstein equation
which is approximately Lorentz invariant by
superposition of quasi-randomly positioned
and quasi-randomly moving $l=0$ defects
(arranged to be nonintersecting, at least initially).
In addition to the expected broadening of light
beams (a Brownian-motion-type effect),  
such a Lorentz-invariant gas of defects may lead to
mass-generation effects~\cite{KlinkhamerQueiruga2017}.

Remark that the $l(w) \ne 0$ stealth-defect solution
of Fig.~4 in Ref.~\citen{KlinkhamerQueiruga2018-stealth}
has a curved spacetime
(parameter choice $\widetilde{\eta} = 1/10$),
which results in some additional bending
of light passing near the defect surface.
In fact, the bending is outwards, as the effective mass
near the defect surface is negative; see the $l(w)$ panel of
Fig.~4 in Ref.~\citen{KlinkhamerQueiruga2018-stealth}
and the definition of $l(w)$ in the sentence below             
\eqref{eq:A1c} in the Appendix of the present article.                               Still, the lensing property is essentially the same as for
the flat-spacetime defect
(see \ref{app:Geodesics-curved-spacetime-stealth-defect}  
for a simplified calculation).

In the lensing argument of Sec.~\ref{sec:Image-formation}
for the $l=0$ flat-spacetime defect,
we considered light rays. But, with the particle--wave duality,
we can also interpret
the three geodesics crossing the defect surface in Fig.~\ref{fig6}
as coherent light emitted from the source $P$
(as mentioned before, the emission is assumed to last for a long time).
At the point $P'$, these coherent-light bundles
have a constant (time-independent) phase difference,
which ​leads 
to stationary interference.
In this sense, our defect resembles not only a
material lens  in Minkowski spacetime
but also some type of interferometer
(the behavior depends primarily on the ratio
of the wavelength $\lambda$ and the defect length scale $b$).

As a final comment, we contrast the lensing
from our hypothetical spacetime defect with standard gravitational
lensing~\cite{Einstein1936,VirbhadraEllis2000,Perlick2004,%
Schneider-etal1992,Wambsganss1998,Dodelson2017}.
Standard gravitational lensing can be interpreted as being due to
the curvature of spacetime resulting from a nonvanishing
matter distribution.
The lensing of Fig.~\ref{fig6} is, however, entirely due to
the nontrivial topology from the defect,
as the spacetime manifold of this particular solution is flat.

\begin{appendix}
\section{Geodesics of a curved-spacetime stealth defect}
\label{app:Geodesics-curved-spacetime-stealth-defect}

In Sec.~\ref{sec:Image-formation},
we have shown that a particular defect in flat spacetime resembles a
material lens in Minkowski spacetime. In this appendix, we will see that the
same resemblance holds for the corresponding defect in curved spacetime.

\subsection{General results}
\label{subapp:General-results}

The general spherically symmetric \emph{Ansatz} for the
metric of a spacetime defect is given by the following line
element~\cite{Klinkhamer2014-prd}:
\bsubeqs\label{eq:A1}
\beqa\label{eq:A1a}
ds^2\,\Big|^\text{(gen.\,sol.)}
&=&-M(w)\,(dt)^2+N(w)\,(dy)^2
+w\, \Big [ (dz)^2+\sin^2z\,(dx)^2 \Big]\,,
\\[2mm]
\label{eq:A1b}
M(w)&\equiv& \big [\mu (w) \big]^2 \,,
\\[2mm]
\label{eq:A1c}
N(w)&\equiv& (1-y_0 ^2/w) \big [\sigma (w) \big]^2 \,,
\eeqa
\esubeqs
with $w$ defined by \eqref{eq:w-def}.
The effective mass parameter $l(w)$
is defined~\cite{KlinkhamerQueiruga2018-antigrav}
by setting $\sigma^2(w) \equiv 1 \big/\big(1-l(w)/\sqrt{w}\big)$.
The functions $\mu (w)$ and $\sigma (w)$ are determined
by the field equations
and the boundary conditions. At this moment, we do not need to know
the explicit form of these functions.

As mentioned in Sec.~\ref{sec:Geodesic equations},
we only need to consider the particle moving in the equatorial plane, $z=\pi/2$.
Then, the nonvanishing Christoffel symbols are
\bsubeqs\label{eq:A2}
\beqa
\christoffel{t}{t}{y}&=&\christoffel{t}{y}{t}=-\frac{1}{2M}\frac{d M}{dy}\,,
\\[2mm]
\christoffel{y}{t}{t}&=&-\frac{1}{2N}\frac{d M}{dy}\,,
\\[2mm]
\christoffel{y}{y}{y}&=&\frac{1}{2N}\frac{d N}{dy}\,,
\\[2mm]
\christoffel{y}{x}{x}&=&-\frac{1}{2N}\frac{d w}{dy}\,,
\\[2mm]
\christoffel{x}{x}{y}&=&\christoffel{x}{y}{x}=
\frac{1}{2w}\frac{d w}{d y}\,.
\eeqa
\esubeqs
With the procedure used in Sec.~\ref{sec:Geodesic equations},
the geodetic equation gives
\bsubeqs\label{eq:A3}
\beqa
\label{eq:A3a}
\frac{dt}{d\lambda}&=&M\,,
\\[2mm]
\label{eq:A3b}
w\, \frac{dx}{d\lambda}&=& \widetilde{J}\,,
\\[2mm]
\label{eq:A3c}
N \left( \frac{dy}{d\lambda}\right)^2 + \frac{\widetilde{J}^{\,2}}{w}-\frac{M^3}{3}
&=& \widetilde{E}\,,
\eeqa
\esubeqs
where $\widetilde{J}$ and $\widetilde{E}$ are real constants and
$\lambda$ is the affine parameter.
By elimination of $\lambda$ from \eqref{eq:A3b} and \eqref{eq:A3c}, we have
\begin{equation}\label{eq:A4}
\frac{\widetilde{J}^{\,2}\,N(w)}{w^2} \left( \frac{dy}{dx}\right)^2
+ \frac{\widetilde{J}^{\,2}}{w}-\frac{[M(w)]^3}{3}= \widetilde{E}\,,
\end{equation}
where the explicit $w$-dependence of $N$ and $M$ has been restored.
With the replacement \eqref{eq:replacement-dy/dx},
condition \eqref{eq:A4} can be written as
\begin{equation}\label{eq:A5}
\frac{\widetilde{J}^{\,2}\,N(w)}{4y^2w^2} \left( \frac{dy^2}{dx}\right)^2
+ \frac{\widetilde{J}^{\,2}}{w}-\frac{[M(w)]^3}{3}= \widetilde{E}\,,
\end{equation}
with the constants $\widetilde{J}$ and $\widetilde{E}$
from \eqref{eq:A3}.

The orbit of a particle moving in the equatorial plane $z=\pi/2$
is described by \eqref{eq:A5}.
Observe that $N(w)$ and $M(w)$ are functions of $w$ and, hence,
functions of $y^2$. If the solution of \eqref{eq:A5} exists,
$x$ must be a function of $y^2$: $x=x(y^2)$.
Recall from \eqref{eq:chart-coord-ranges} that
the chart-2 coordinate ranges are given by
\begin{equation}\label{eq:A6}
x \in (0,\pi)\,, \ \
y\in (-\infty,\infty)\,, \ \
z \in (0,\pi)\,.
\end{equation}
For a particular solution $x=x(y^2)$ in the $z=\pi/2$ plane of
the chart-2 domain, there are then two branches:
one branch with $y \ge 0$ and the other one with $y \leq 0$
(note that the point $y=0$ has been included for both branches,
as was done in
Sec.~\ref{sec:Nonradial-geodesics}).
To be specific, the lines which correspond to these two branches
of the solution are symmetrical about the ``center" of the defect surface.
If the orbit of a given particle
does not cross the defect surface, then this orbit
is usually described by only one of these two branches.
But, if the particle crosses the defect surface,
then we argue that the ingoing and outgoing lines are given by two different branches.
Remark that, in flat spacetime, this argument is consistent with the two conditions
for the existence of a unique
outgoing line as discussed in Sec.~\ref{subsec:Geodesics-crossing-the-defect-surface}.

Based on above points, a defect in a curved spacetime
resembles a material lens and has the same properties as discussed
in Sec.~\ref{sec:Image-formation} for the flat-spacetime case.
Still, there is one exception: a black hole
may occur for this defect spacetime~\cite{Klinkhamer2014-mpla}.
Then, the metric \eqref{eq:A1} is not globally regular
and \eqref{eq:A5} cannot properly describe the orbit of the
particle reaching the defect surface. In fact, the particle
will be confined within the black-hole horizon once it crosses the horizon
(appropriate coordinates would, for example,  be
the Painlev\'{e}--Gullstrand-type coordinates
of App.~C in Ref.~\citen{Klinkhamer2014-mpla}).

\subsection{Explicit calculation}
\label{subapp:Explicit-calculation}

The numerical stealth-defect solution from Fig.~4 of
Ref.~\citen{KlinkhamerQueiruga2018-stealth} has
metric functions
$\sigma(w)$ and $\mu(w)$ in \eqref{eq:A1}
with approximately the following form:
\bsubeqs\label{eq:A7}
\beqa
\sigma(w)&=&1-\frac{1}{2w}\,,
\\[2mm]
\mu(w)&=&1\,,
\eeqa
\esubeqs
for $y_0=1$ (giving $w \equiv 1+ y^2$).
We will now obtain the analytic solutions
of \eqref{eq:A3} and \eqref{eq:A5} from the
explicit choice of functions in \eqref{eq:A7}.

\begin{figure}[t]
  \centering
  \includegraphics[width=0.55\textwidth]{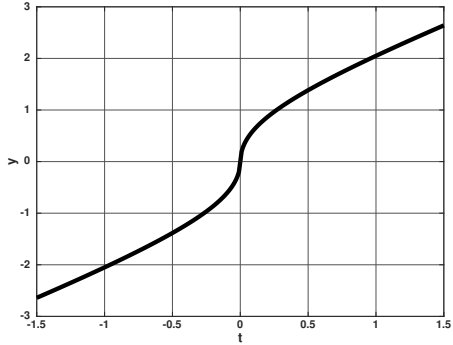} 
  \vspace*{0mm}
  \caption{Null radial geodesic
  for the stealth defect \eqref{eq:A1} with metric functions \eqref{eq:A7},
  using dimensionless coordinates $y$ and $t$.}
\label{fig8}
\end{figure}

For the radial geodesic ($\widetilde{J}=0)$,
the general solutions of \eqref{eq:A3c} are
\bsubeqs\label{eq:A8}
\beqa
\frac{2 w (2 w+1)\, \sqrt{(-2 w+1)^2/w^3}}{2 w-1}&=&
+4\,t \, \sqrt{\widetilde{E}+1/3}+\widetilde{C}_1 \,,
\\[2mm]
\frac{2 w (2 w+1)\, \sqrt{(-2 w+1)^2/w^3}}{2 w-1}&=&
-4\,t \, \sqrt{\widetilde{E}+1/3}+\widetilde{C}_2 \,,
\eeqa
\esubeqs
where $\widetilde{C}_1$ and  $\widetilde{C}_2$ are real constants.
An example of a null radial geodesic is shown in Fig.~\ref{fig8}.

For a nonradial geodesic, the solutions of \eqref{eq:A5} are
\bsubeqs\label{eq:A9}
\beqa
\pm x &=&  \frac{1}{4}\left( (4-D)\arctan
(\sqrt{D\,w-1}) -\frac{\sqrt{D\,w-1}}{w}\right)+\widetilde{x}_4 \,,
\eeqa
with the definition
\beqa
D  &\equiv& \frac{\widetilde{E}+1/3}{\widetilde{J}^{\,2}}
\eeqa
\esubeqs
and a real constant $\widetilde{x}_4$.

For geodesics that do not cross the defect surface,
we can, just as in
Sec.~\ref{subsec:Geodesics-not-crossing-the-defect-surface},
calculate the change in $x$,
\begin{equation}\label{eq:A10}
\Delta x\equiv |x(y_1)-x(\infty)|=
\frac{\pi}{2}\left( 1-\frac{1/4}{1 +y_1 ^2}\right) \,,
\end{equation}
where $y_1$ corresponds to the point on the line closest to the defect surface.
For small $y_1$ (i.e., the line coming close to the defect surface),
\eqref{eq:A10} shows that the line is bent away from the defect surface.
This agrees with
the fact that the effective mass near the defect surface is negative;
see the $l(w)$ panel in
Fig.~4 of Ref.~\citen{KlinkhamerQueiruga2018-stealth}.

\begin{figure}[]
  \centering
  \includegraphics[width=1\textwidth]{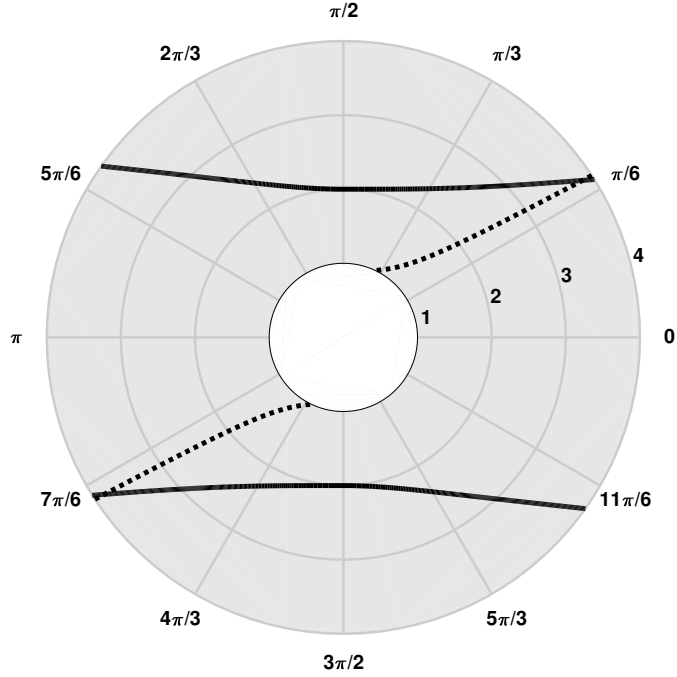}
  \caption{
  Geodesics in polar-type coordinates $(\sqrt{w},\, \phi)$,   
  where the geodesics are given by \eqref{eq:A12}.
  With the chart-2 coordinates $x$ and $y$,
  the azimuthal angle $\phi$ is defined by
  $\phi=x$ if $y>0$ and $\phi=x+\pi$ if $y<0$.
  The defect surface is given by the circle $w=1$
  and part of the 3-dimensional space manifold  \eqref{eq:A1},
  with metric functions \eqref{eq:A7}, is indicated by the shaded area.
  The solid lines have constants
  $D=0.25$ and $x_4=\pi/2$ and the dotted-line segments have constants
  $D=1.25$ and $x_4=\pi/2$.
  The points on the solid lines which are closest
  to the defect surface have polar-type coordinates   
  $(2,\, \pi/2)$ and $(2,\, 3\pi/2)$, corresponding to the original
  coordinates $(y,\, x)=(\pm \sqrt{3},\, \pi/2)$.   }
  \label{fig9}
\end{figure}

\begin{figure}[]
\includegraphics[width=1\textwidth]{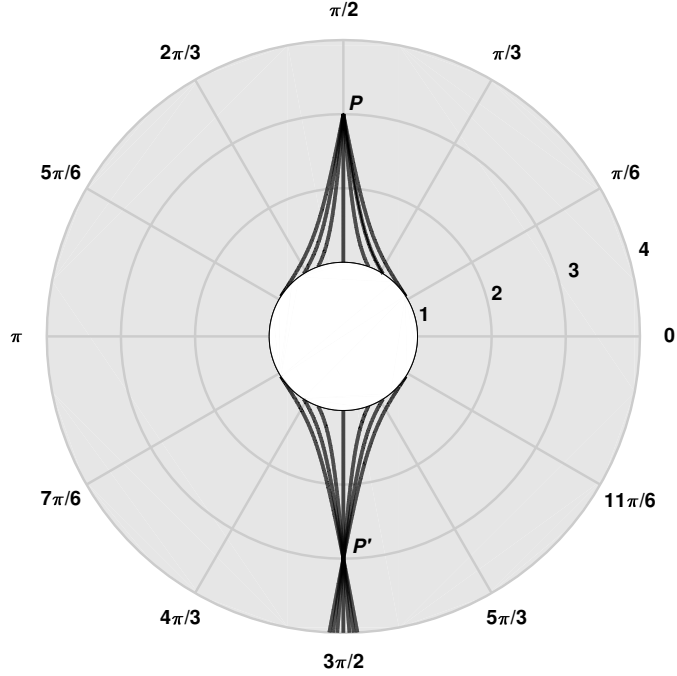}
\caption{
  Family of geodesics in polar-type coordinates $(\sqrt{w},\, \phi)$,  
  where the geodesics are given by \eqref{eq:A11}
  with plus signs on the left-hand sides.
  The defect surface is given by the circle $w=1$.
  The parameters of the six curved geodesics in the upper half-plane
  are, from left to right, ($D=1,~x_4=2.57258$), ($D=1.2,~x_4=2.5408766$),
  ($D=1.5,~x_4=2.4784766$), ($D=1.5,~x_5=\pi-2.4784766$),
  ($D=1.2,~x_5=\pi-2.5408766$), and ($D=1,~x_5=\pi-2.57258$).
  In terms of the original $(y,\, x)$ coordinates,
  the focal points $P$ and $P'$
  are given by $(y,\, x)_{P}=(\sqrt{8},\, \pi/2)$
  and $(y,\, x)_{P'}=(-\sqrt{8},\, \pi/2$).
  }
  \label{fig10}
\end{figure}

Note that \eqref{eq:A9} can be rewritten in the following way:
\bsubeqs\label{eq:A11}
\beqa
\label{eq:A11a}
\pm\frac{1}{\sqrt{D}}&=&\sqrt{w}\,
\cos \left[ \frac{4(x-x_{4})+\sqrt{D\,w-1}/w}{4-D}\right] \,,
\\[2mm]
\label{eq:A11b}
\pm\frac{1}{\sqrt{D}}&=&\sqrt{w}\,
\cos \left[ \frac{4(-x+x_{5})+\sqrt{D\,w-1}/w}{4-D}\right] \,,
\eeqa
\esubeqs
with real constants $x_4$  and $x_5$.
As a concrete example, we first consider
the solution corresponding to the upper sign on the left-hand side of
\eqref{eq:A11a}, that is,
\begin{equation}\label{eq:A12}
\frac{1}{\sqrt{D}}=
\sqrt{w}\,\cos \left[ \frac{4(x-x_{4})+\sqrt{D\,w-1}/w}{4-D}\right].
\end{equation}
For given values of $D$ and $x_4$,
the solution \eqref{eq:A12} has,  in general, two branches:
one branch lies in the upper half-plane ($y>0$)
and the other in the lower half-plane ($y<0$).
The solid lines in Fig.~\ref{fig9}
correspond to the orbits of two different particles, while
the dotted line corresponds to the orbit of a third particle.
Even though the points on the solid lines which are closest
to the defect surface have $x=\pi/2$,
these solid lines are not symmetrical about the line $x=\pi/2$
for $w \sim 2$,
as can be verified in \eqref{eq:A12} with $x'=\pi-x$ and $w(x') \neq w(x)$.

In Fig.~\ref{fig10},
finally, we give a family of geodesics in order to illustrate
the lensing property of the curved-spacetime defect
(cf.~Fig.~\ref{fig6} for the lensing of the flat-spacetime defect).
Apparently, the spherical symmetry of the metric
is the crucial input for the lensing behavior.

\end{appendix}

\section*{Acknowledgments}
\vspace*{-0mm}
\noindent
We thank J.M.~Queiruga and the referees for useful comments.
The work of Z.L.W.
is supported by the China Scholarship Council.


\end{document}